# Impact of Ni doping on critical parameters of PdTe type-II BCS superconductor

*Rajveer Jha[1], Reena Goyal[1,2], Brajesh Tiwari[3] and V.P.S. Awana[1,2,*]*

[1]*CSIR-National Physical Laboratory, Dr. K. S. Krishnan Marg, New Delhi-110012, India*
[2] *Academy of Scientific and Innovation Research, NPL, New Delhi-110012*
[3]*Sardar Vallabhbhai National Institute of Technology, Surat-395007, Gujarat, India*

**Abstract**

We report the impact of Ni doping on superconductivity of PdTe superconductor. The superconducting parameters like critical temperature ($T_c$), upper critical field ($H_{c2}$) and normalized specific-heat jump ($\Delta C/\gamma T_c$) are reported for Ni doped $Pd_{1-x}Ni_xTe$. The samples of series $Pd_{1-x}Ni_xTe$ with nominal compositions x=0, .01, 0.05, 0.07, 0.1, 0.15, 0.2, 0.3 and 1.0 are synthesized via solid state reaction route. All the studied samples of series $Pd_{1-x}Ni_xTe$ (x = 0.0 to 1.0) are crystallized in hexagonal crystal structure within the space group *P6₃/mmc*. Unit cell volume shrinks almost linearly upon Ni doping in $Pd_{1-x}Ni_xTe$. The normal state residual resistivity increases with Ni substitution on Pd site. Both the electrical resistivity and magnetic measurements revealed that $T_c$ decreases with increase of Ni concentration in $Pd_{1-x}Ni_xTe$ and is not observed down to 2K for x=0.30 i.e., 30% of Ni doping at Pd site. Interestingly, this is unusual for magnetic Ni doping in a known type-II BCS type superconductor. Magnetic Ni must suppress the superconductivity much faster. Interestingly, the isothermal magnetization measurements for NiTe revealed that Ni is non-magnetic in $Pd_{1-x}Ni_xTe$ structure and hence the $T_c$ depression is mainly due to disorder. The magneto-transport measurements revealed that flux is better pinned for 20% Ni doped PdTe as compared to other compositions of $Pd_{1-x}Ni_xTe$. The magnetic field dependence of specific heat of $Pd_{1-x}Ni_xTe$ for x=0.01 was studied and the estimated value of the normalized specific-heat jump, $\Delta C/\gamma T_c$, is found to be 1.42, which is under BCS weak-coupling limit. Summarily, we report the impact of Ni doping in $Pd_{1-x}Ni_xTe$ superconductor and conclude that Ni substitutes at Pd site, suppress superconductivity moderately and is of non magnetic nature in this system. To best of our knowledge this is the first study on Ni substitution in PdTe superconductor.

[*]**Corresponding Author**
Dr. V. P. S. Awana:  E-mail: awana@mail.npindia.org
Ph. +91-11-45609357, Fax-+91-11-45609310
Homepage awanavps.webs.com

**Introduction**

The discovery of new superconductors always attracted enormous interest from both experimental and theoretical condensed matter physics community. For the meantime, the already discovered superconductors keep on motivating to understand the underlying physics of them. For example, the impact of magnetic impurities in known superconductors has been of great interest for a long time. Adding magnetic impurity to a superconductor does provide an avenue to induce interesting new class of matter and to probe the fundamental mechanism of exotic ground states of strongly correlated electron materials [1-4]. For example the magnetic Mn impurities in $Ba_{0.5}K_{0.5}Fe_2As_2$ and $Ba(Fe_{1-y}Co_y)_2As_2$ systems showed that $T_c$ is strongly suppressed [5-7], while $T_c$ is nearly unchanged in Mn-doped $FeSe_{0.5}Te_{0.5}$ superconductors [8]. On the other hand, non magnetic Zn doped $BaFe_{1.89-2x}Zn_{2x}Co_{0.11}As_2$ compounds showed that the $T_c$ decreases rapidly with increasing Zn doping level [9], but the superconducting state is quite robust for $Fe_{1-y}Zn_ySe_{0.3}Te_{0.7}$ compound [10]. In case of high $T_c$ Cuprates significant decrease in $T_c$ was observed with Cu site Zn doping [11-13]. Studying the impact of both magnetic/non magnetic impurities on known superconductors had been of prime interest for over the years [1-13].

Recently, a new superconductor PdTe came into existence in both single crystal and bulk polycrystalline form [14, 15]. This work motivated us to study the effect of magnetic Nickel on superconductivity of bulk polycrystalline PdTe. The effects of magnetic impurities and the possibility of magnetic ordering in BCS type conventional PdTe superconductor could provide better understanding of the phenomenon. Generally, it has been believed that the conduction electrons cannot be ordered both magnetically and superconducting due to strong spin scattering [16, 17]. On the other hand Cooper pairs are formed in Cuprates and Iron based compounds possibly through spin fluctuations and superconductivity occurs after suppressing the magnetic ordering by chemical doping or the application of hydrostatic pressure [18-20]. The electron-phonon coupling as proposed in BCS theory failed to explain the superconductivity in Cuprate and Iron based materials [21]. The superconductivity in high $T_c$ Cuprates is induced from electronic charge carriers doping in antiferromagnetic Mott insulating phase [21–23]. There is a hypothetical possibility of the magnetic excitations being replacing phonons in exotic high $T_c$ superconductors [23]. On the other hand, there are some examples for the coexistence of superconductivity with either ferromagnetic or anti-ferromagnetic ordering like $UGe_2$, URhGe, UCoGe, $MgCNi_3$ and $RuSr_2GdCu_2O_8$ etc [24-28]. As far as the coexistence of superconductivity and magnetism is concerned, there is so far no one theory detailing the interaction between superconducting and magnetic order parameters. In some experimental reports, it has been suggested that the $T_c$ decreases linearly with increasing magnetic impurity concentration in superconducting systems [6-9]. The decrease in $T_c$ of bulk lanthanum by rare-earth impurities depends on the spin of the impurity atoms rather than on their magnetic moment, which has been reported by Matthias in a detailed study [29-33].

Keeping in view, the wider interest of the condensed matter physics community in doping of magnetic impurities in known superconductors, here, we report on synthesis and characterisation of series samples of $Pd_{1-x}Ni_xTe$ with x=0, .01, 0.05, 0.07, 0.1, 0.15, 0.2, 0.3 and 1.0. The X-ray diffraction (XRD) measurements revealed that the Ni gets substituted at Pd site in the parent hexagonal phase (space group $P6_3/mmc$) of PdTe. The superconducting transition temperature $T_c$ of $Pd_{1-x}Ni_xTe$ is studied by resistivity measurements using QD PPMS down to 2 K under different magnetic fields. The Heat capacity measurements for $Pd_{0.99}Ni_{0.01}Te$ are also presented and analysed. Ni doping in $Pd_{1-x}Ni_xTe$ decreases superconductivity moderately, and the reason behind is that Ni is of non magnetic nature in PdTe. The Ni(3d) and Te(sp) orbital possible strong hybridisation might be the reason behind non magnetic nature of Ni in $Pd_{1-x}Ni_xTe$. Detailed first principal density functional calculations along with photo electron spectroscopy studies could shed more light on the cause behind non magnetic nature of Ni in $Pd_{1-x}Ni_xTe$ system. Interestingly, to best of our knowledge this is the first study on Ni substitution at Pd site in PdTe superconductor.

**Experimental**

The bulk Polycrystalline $Pd_{1-x}Ni_xTe$ (x=0, .01, 0.05, 0.07, 0.1, 0.15, 0.2, 0.3 and 1.0) samples were synthesized by the solid state route via vacuum encapsulations. The essential elements Pd (99.9%-3N), Te (99.99%-4N) and Ni (99.99%-4N) from Sigma Aldrich are mixed in a stoichiometric ratio in an argon filled glove box and then pelletized by applying uniaxial stress of 100kg cm$^{-2}$. The pellets sealed in an evacuated (<10$^{-3}$ Torr) quartz tubes were kept in a furnace for heating at 750°C at a rate of 2°C min$^{-1}$ for 24hours. The obtained samples were dense and shiny black. For different physical property measurements, the samples were broken into desired pieces. For the structural analysis we have used a Rigaku x-ray diffractometer with Cu Kα radiation of 1.5418Å. Electrical, magnetic and heat capacity ($C_P$) measurements were carried out on a Quantum Design (QD) Physical Property Measurement System (PPMS-14Tesla)-down to 2K.

**Results and Discussion**

Powder x-ray diffraction patterns are recorded for as prepared samples at room temperature. Figure 1 shows the observed and Reitveld fitted room temperature x--ray diffraction pattern for the $Pd_{1-x}Ni_xTe$ (x=0, .01, 0.05, 0.07, 0.1, 0.15, 0.2, 0.3 and 1.0) samples. All the samples are well fitted with the space group $P6_3/mmc$, suggesting complete solubility of Ni in PdTe. It can be clearly seen from Fig.1 that (1 0 0) crystallographic plane around $2\theta=24.82^o$ is being suppressed with increasing concentration of Ni and for x=0.2 or above the same disappears completely. On the other hand the (0 0 2) plane at $2\theta=33.07^o$ appears only above x=0.1 in the main phase of PdTe compound, and is clearly absent for below x=0.15 Ni doped samples. It is not clear to us at this juncture that as if some structural transformation takes place in $Pd_{1-x}Ni_xTe$ system at x =0.20. As far as the lattice parameters are concerned the same are; $a=b=4.1533(2)$Å and $c=5.6733(5)$Å for PdTe and $a=b=3.941(2)$Å and $c =5.3632(5)$ Å for NiTe within the $P6_3/mmc$ space group. The reitveld refined lattice parameters *a, b, c* and volume (*V*) are found to be consistently decreasing with increasing Ni content in $Pd_{1-x}Ni_xTe$, see Figure 2. Almost linear shrinkage of the unit cell volume of $Pd_{1-x}Ni_xTe$ with x indicates successful substitution of Pd by Ni in PdTe. It is worth noting that both the lattice parameters *a* and *c* (Figure 3b) decrease simultaneously suggesting that chemical pressure being exerted by Pd site Ni substitution is isotropic in nature. There is clear indication of chemical pressure on the unit cell of PdTe within same hexagonal crystal structure with Pd site Ni substitution. This pressure may play an important role on the superconductivity of parent PdTe compound. For Pd doped FeTe compound, it has been reported that the negative chemical pressure as well as doping induced structural phase transition occurs from tetragonal to hexagonal phase [34].

Figure 3 represents the ac susceptibility for the all superconducting $Pd_{1-x}Ni_xTe$ (x=0, 0.01, 0.05, 0.07, 0.1, 0.15 and 0.2) samples. Both the real (M') and imaginary (M'') parts of the ac magnetic susceptibility measurements are carried out at an amplitude of 10Oe and frequency 333Hz down to 2K. The M' showed a sharp transition to diamagnetism ($T_c$) at around 4.5K, confirming the bulk superconductivity in pristine PdTe sample. In contrast, M" exhibited a sharp single, positive peak around the same temperature, indicating good coupling of superconducting grains in PdTe superconductor. With increasing Ni concentration in $Pd_{1-x}Ni_xTe$, the $T_c$ shifts monotonically to lower temperatures from 4.5K (x=0.0) to just above 2.5K for x= 0.20 sample. The Ni doped PdTe samples are superconducting up to the doping level of x=0.2 and for the higher value of x the same become non superconducting at least down to 2K.

Figure 4 shows the Isothermal magnetization curves at 2K to estimate the lower critical field ($H_{c1}$) values of superconducting $Pd_{1-x}Ni_xTe$ samples. With increasing magnetic field from zero, the absolute value of magnetization increases linearly up to $H_{c1}$, signifying the complete diamagnetic character. For pure PdTe the absolute value of demagnetization starts decreasing above the magnetic field $H_{c1}$ (lower critical field), reaches zero and becomes positive above an applied field of 1kOe at 2K upper critical field $H_{c2}$. Clearly the magnetization curves confirm the PdTe to be a type-II superconductor. The estimated $H_{c1}$ values are 200Oe, 160Oe, 51Oe, 41Oe, and 32Oe for $Pd_{1-x}Ni_xTe$ (x=0, 0.01, 0.05, 0.07 and 0.1) respectively. Clearly the $H_{c1}$ of $Pd_{1-x}Ni_xTe$ decreases with Ni doping. Inset of Figure 4 shows the isothermal magnetization (MH) plot for NiTe at room temperature (300K) in applied fields of up to ±5000Oe. Interestingly, though the MH indicates towards ferromagnetism, the absolute value of magnetic susceptibility is too small to account for the magnetic moment of Ni spins. As can be seen from inset of Fig. 4, the seemingly saturation moment is of the order of $10^{-5}\mu_B$/Ni. It is clear from inset of Figure 4 that Ni is of non magnetic nature in NiTe. The non magnetic nature of Ni has earlier been indicated in NiTe nanowires [35].

Temperature dependence of electrical resistivity for $Pd_{1-x}Ni_xTe$ (x=0, .01, 0.05, 0.07, 0.1, 0.15, 0.2, 0.3 and 1.0) samples in temperature range 300-2K is shown in Figure 5a. Normal state electrical resistivity of all the samples exhibits the metallic behaviour. The normal state electrical resistivity say above 200K decreases with increasing Ni content up to x=0.1 and then increases for higher Ni content and the same is highest for the NiTe sample. Figure 5b is the ρ(T) graph of Ni doped superconducting compounds in the temperature range 6-2K. The normal state resistivity (near to $T_c$ onset) increases, the superconducting transition temperature ($T_c$) decreases with increasing Ni substitution at Pd site. This trend is shown in Figure 6 in terms of $T_c^{onset}$ vs Ni content plot for $Pd_{1-x}Ni_xTe$ superconducting samples. Here one can see that the $T_c^{onset}$ is nearly unchanged for x=0.01 sample and the same decreases rapidly for higher Ni content samples. Both 30% Ni doped and NiTe samples are though non superconducting down to 2K but having metallic behaviour of normal state electrical resistivity. Figure 7 shows the fitted electrical resistivity equation $\rho=\rho_o+ AT^2$, where $\rho_o$ is the residual resistivity and A is the slope of the graph. Fitting of the ρ(T) graph is shown as red line in the temperature range of 5-50K, see inset of Figure 7. The obtained values (Table 1) of residual resistivity clearly increase with increasing concentration of Ni in $Pd_{1-x}Ni_xTe$, suggesting increased disorder/defects in Ni doped samples. $\rho_0$ value is highest for NiTe sample, which is though metallic but non superconducting as well. A similar trend of increasing $\rho_0$ with 3d metal doping is reported earlier for the $Na(Fe_{0.97-x}Co_{0.3}T_x)As$ (T = Cu, Mn) and Co-doped $Fe_{1+y}Te_{0.6}Se_{0.4}$ superconducting systems [36,37]. In our case the residual resistivity increases monotonically up to the doping level of 20%, and later shoots up as the superconductivity disappears. The suppression of $T_c$ in $Pd_{1-x}Ni_xTe$ system may result from the change of the charge carrier density along with the impurity scattering. Magnetic pair braking is ruled out because Ni is non magnetic in $Pd_{1-x}Ni_xTe$ as revealed from our magnetization measurements.

Figure 8 (a-f) demonstrate the temperature dependence of electrical resistivity under various magnetic field in the temperature range 2-6K for superconducting $Pd_{1-x}Ni_xTe$ (x=0, .01, 0.05, 0.07, 0.1, 0.15 and 0.2) samples. The $T_c^{onset}$ and $T_c^{(\rho=0)}$ decreases with increasing magnetic field, i.e., a typical type-II superconducting behaviour has been observed for all the samples. Figure 8h shows the upper critical field $H_{c2}$ corresponding to the temperatures where the resistivity drops to 90% of the normal state resistivity. It is well known phenomena that the magnetic field interact with conduction electrons in nonmagnetic superconductors, which follow basically two different mechanisms. One is orbital pair breaking due to interaction of field with the orbital motion of electron and other is interaction of field with the electron spin i.e. Pauli paramagnetic limiting effects. At low fields the orbital pair breaking mechanism is dominating and for very high fields the Pauli paramagnetic effect limits the upper critical field. The $H_{c2}(0)$ is estimated by using the conventional one-band Werthamer–Helfand–Hohenberg (WHH) equation, i.e., $H_{c2}(0)=-0.693T_c(dH_{c2}/dT)_{T=Tc}$. The solid lines are the ones being extrapolated to T = 0 K, for 90% $\rho_n$ criteria of $\rho(T)H$ curve for $Pd_{1-x}Ni_xTe$ samples. The estimated $H_{c2}(0)$ values are 2.6kOe, 2.3kOe, 2.4kOe, 2.6kOe, 2.61kOe, 2.66kOe and 3kOe for $Pd_{1-x}Ni_xTe$ (x=0, .01, 0.05, 0.07, 0.1, 0.15 and 0.2) samples. The $H_{c2}(0)$ value for 20% Ni doped PdTe is significantly higher than the pristine sample, while the $T_c$ (3K) is lower than the $T_c$(4.5K) of PdTe sample. This suggests strong pinning due to impurity scattering effect in Ni doped PdTe compound [38]. The estimated $H_{c2}(0)$ values for all the samples are within the Pauli paramagnetic limit ($\mu_oH_p=1.84T_c$), which can be considered as an evidence for the conventional nature of superconductivity in PdTe.

Low temperature heat capacity under various magnetic fields for $Pd_{1.99}Ni_{0.01}Te$ is shown in Figure 9a. At zero fields, a sharp kink at temperature 4.5K is observed due to the superconducting transition, which decreases to low temperature with increasing magnetic field. The low temperature specific heat data i.e., below $T_c$ gives important information about the superconducting energy gap structure. Inset of Figure 9a shows the specific heat divided by temperature ($C_p/T$) as a function of $T^2$. The superconductivity anomaly/kink is completely suppressed and not seen down to 2K at applied magnetic field of 1.5kOe. Thus obtained electronic specific heat was fitted to the expression $C_p(T) = \gamma T + \beta T^3 + \delta T^5$ and through the best fitting the coefficients obtained are $\gamma = 7.42$ mJ mol$^{-1}$ K$^{-2}$, $\beta = 0.8001$ mJ mol$^{-1}$ K$^{-4}$ and $\delta = 0.0019$ mJ mol$^{-1}$ K$^{-6}$. The Debye temperature ($\theta_D$) is 229.9K, which is calculated by using the relation $\theta_D = (234zR/\beta)^{1/3}$, where R is the Rydberg constant, i.e. 8.314 J mol$^{-1}$ K$^{-1}$ and z is the number of atoms in the Ni doped PdTe unit cell. The Kadowaki–Woods ratio $A/\gamma^2$ is $8.7 \times 10^{-5}$ µΩ cm mol$^2$ K$^2$ J$^{-2}$, where A is evaluated by fitting of temperature dependent resistivity in previous section. Interestingly, the value of Kadowaki–Woods ratio for Ni-1% doped PdTe sample is in good agreement with transition metals system [39]. A clear jump appears in Ce/T at a temperature of 4.5 K see Figure 9b. From inset of Figure 9b it can be clearly seen that the magnitude of the jump (ΔC) at T = $T_c$ is 10.59 mJ/molK$^2$, and the value of the normalized specific-heat jump, (ΔC/γ$T_c$) is 1.42. This value is nearly equal to the simple BCS weak-coupling limit, i.e., 1.43. In the previous report our result showed that the PdTe is weak coupled superconductor and the ΔC/γ$T_c$, is 1.33, which is slightly less than the simple BCS weak-coupling limit [15]. On the other hand the Karki et al obtained 1.67, which is slightly larger than BCS weak-coupling regime suggesting that the PdTe is strongly coupled superconductor [14]. It has also been suggested that the slightly larger ΔC/γ$T_c$ value could be due to small amount of disordered phase in the sample [14,15]. In present case of $Pd_{1.99}Ni_{0.01}Te$ along with our earlier report on PdTe, we found that superconductivity of these compounds is simply with in BCS coupling limit.

**Conclusion**

In summary, we successfully synthesized Ni doped PdTe compounds, the XRD pattern for the $Pd_{1-x}Ni_xTe$ (x=0, .01, 0.05, 0.07, 0.1, 0.15, 0.2, 0.3 and 1). Superconductivity ($T_c$) decreases with increase in Ni content and is completely disappeared at above 20% Ni doping. Interestingly, Ni is found to be of non magnetic nature in $Pd_{1-x}Ni_xTe$, and hence the $T_c$ depression is mainly due to disorder alone. The $H_{c2}(0)$ value for 20% Ni doped PdTe is significantly higher than the pristine PdTe sample, suggesting possible pinning. The value of the normalized specific-heat jump ($\Delta C/\gamma T_c$) of 1.42 is estimated from the analysis of specific heat data of $Pd_{1.99}Ni_{0.01}Te$, suggesting a simple BCS weak-coupling limit. This is first study on Ni doping in PdTe superconductor.

**Acknowledgement**

The authors would like to thank the Director of NPL-CSIR India for his passionate curiosity in the present work. This work is financially supported by a DAE-SRC outstanding investigator award scheme on the search for new superconductors. RG thanks UGC, India, for providing her research fellowship.

Table 1: Normal state resistivity fitted parameters evaluated from equation $\rho=\rho_0+AT^2$

| $Pd_{1-x}Ni_xTe$ | $\rho_0(\mu\ \Omega\text{-cm})$ | $A(\mu\ \Omega\ cm–K^{-2})$ |
|---|---|---|
| x=0 | 5.341 | $6.10038\times10^{-9}$ |
| x=0.01 | 7.683 | $4.82482\times10^{-9}$ |
| x=0.05 | 15.694 | $4.96768\times10^{-9}$ |
| x=0.07 | 17.024 | $3.93908\times10^{-9}$ |
| x=0.1 | 25.819 | $4.09503\times10^{-9}$ |
| x=0.15 | 28.573 | $3.03277\times10^{-9}$ |
| x=0.2 | 43.510 | $3.74791\times10^{-9}$ |
| x=0.3 | 83.347 | $5.26336\times10^{-9}$ |
| x=1 | 242.42 | $6.93269\times10^{-10}$ |

**Figure caption**

**Figures 1:** Experimental (red open circles) and reitveld refined (black solid line) room temperature x-ray diffraction patterns of $Pd_{1-x}Ni_xTe$ (x=0, .01, 0.05, 0.07, 0.1, 0.15, 0.2, 0.3 and 1) compounds. The bottom (blue) lines correspond to the difference between the experimental and calculated data.

**Figures 2:** Nominal x dependence reitveld fitted cell parameters $a(Å)$, $c(Å)$ and $V(Å^3)$ for $Pd_{1-x}Ni_xTe$ samples.

**Figures 3:** Temperature dependence of ac susceptibility for superconducting $Pd_{1-x}Ni_xTe$ (x=0, .01, 0.05, 0.07, 0.1, 0.15, 0.2 and 0.3) compounds.

**Figures 4:** Isothermal magnetization vs dc magnetic field in superconducting state at 2K for $Pd_{1-x}Ni_xTe$ (x=0, .01, 0.05, 0.07 and 0.1) compounds. Inset is Isothermal magnetization curve for NiTe compound at room temperature.

**Figures 5:** Temperature dependence of electrical resistivity of $Pd_{1-x}Ni_xTe$ (x=0, .01, 0.05, 0.07, 0.1, 0.15, 0.2, 0.3 and 1) compounds (a) in the temperature range 300-2K and (b) Zoomed part of the same in the superconducting region 6-2K.

**Figures 6:** Nominal x dependence of $T_c^{onset}$ (K) of superconducting $Pd_{1-x}Ni_xTe$ samples.

**Figures 7:** Nominal x dependence of residual Resistivity ($\rho_0$) of $Pd_{1-x}Ni_xTe$ samples. Inset shows the method of $\rho(T)$ curve fitting in the relation $\rho=\rho_o + AT^2$.

**Figures 8:** (a-g) temperature dependence of electrical resistivity under various magnetic fields of superconducting $Pd_{1-x}Ni_xTe$ (x=0, .01, 0.05, 0.07, 0.1, 0.15 and 0.2) compounds. (h) Upper critical field ($H_{c2}$) as a function of temperature solid lines is linearly extrapolation of experimental data.

**Figures 9:** (a) Temperature dependence of heat capacity ($C_p$) under various magnetic fields of superconducting $Pd_{0.99}Ni_{0.01}Te$ compounds. Inset is $C_p/T$ vs $T^2$ at different fields (b) Specific heat Cp of $Pd_{0.99}Ni_{0.01}Te$ compound at zero and 1.5kOe field. The solid red line is the fit to the relation $C_p(T) = \gamma T + \beta T^3 + \delta T^5$. Inset is change of specific heat $C_e/T$ as a function of temperature.

Fig. 1:

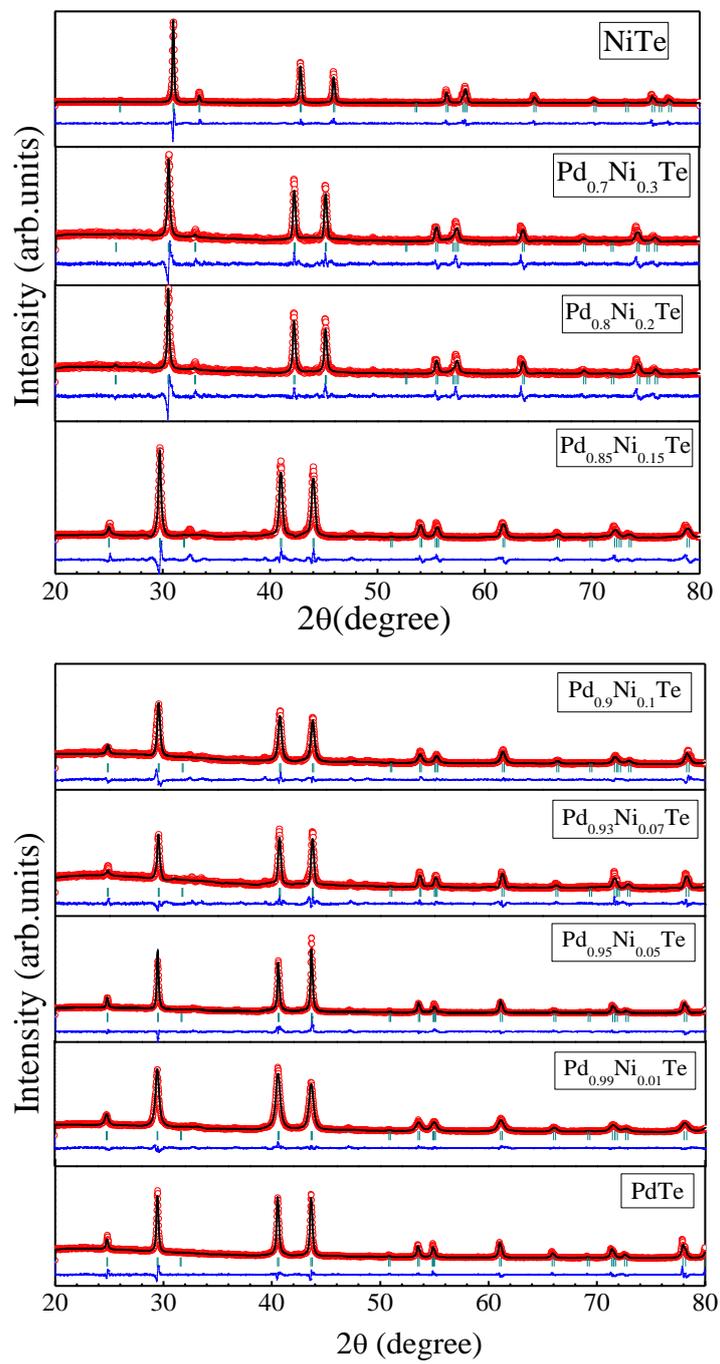

Fig. 2:

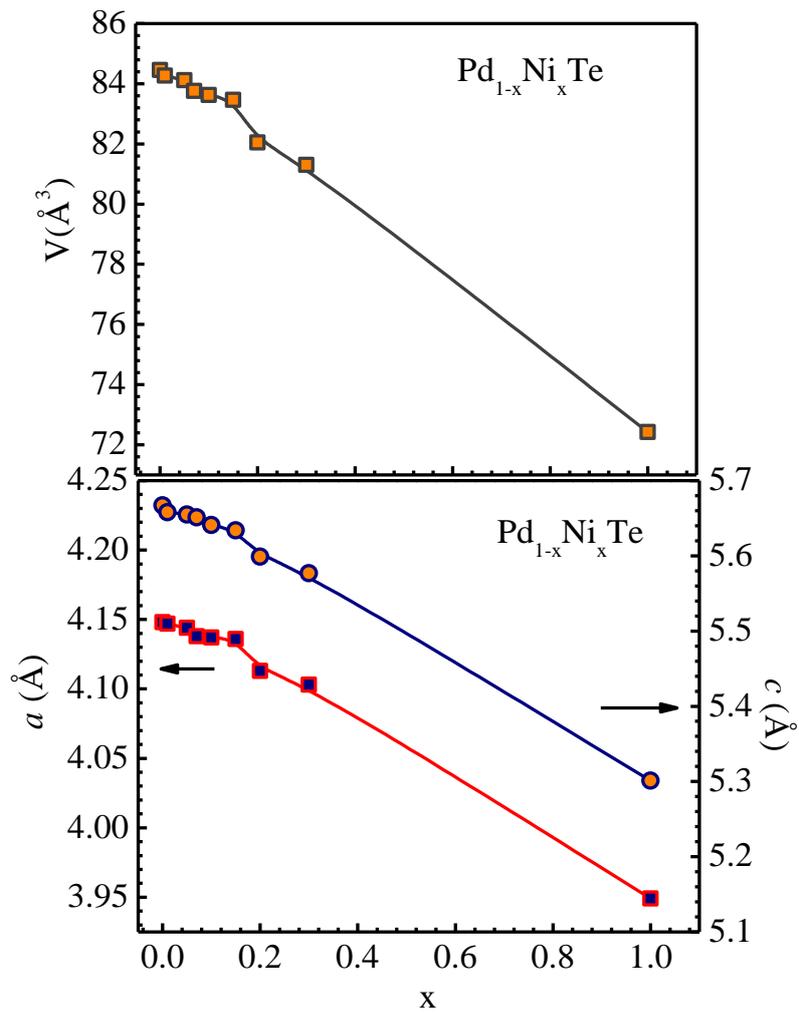

Fig. 3:

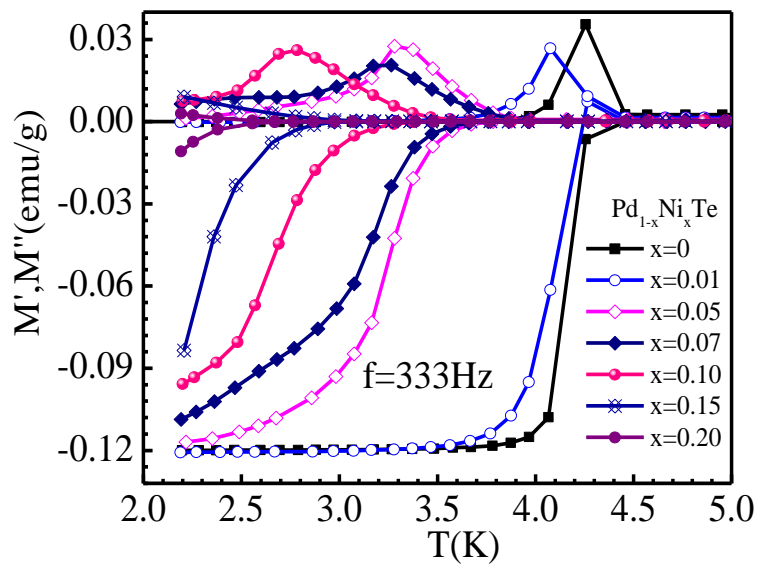

Fig. 4:

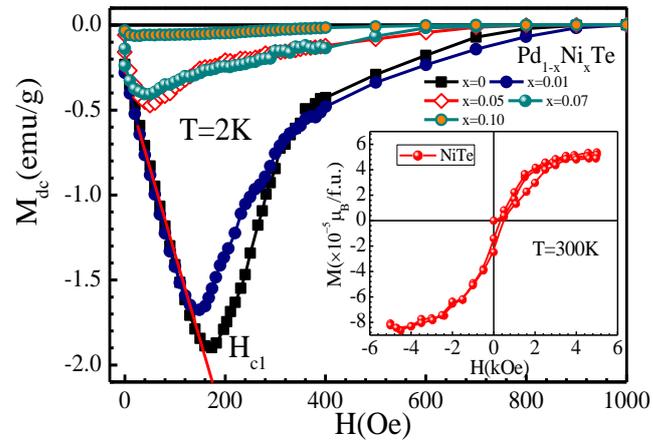

Fig.5:

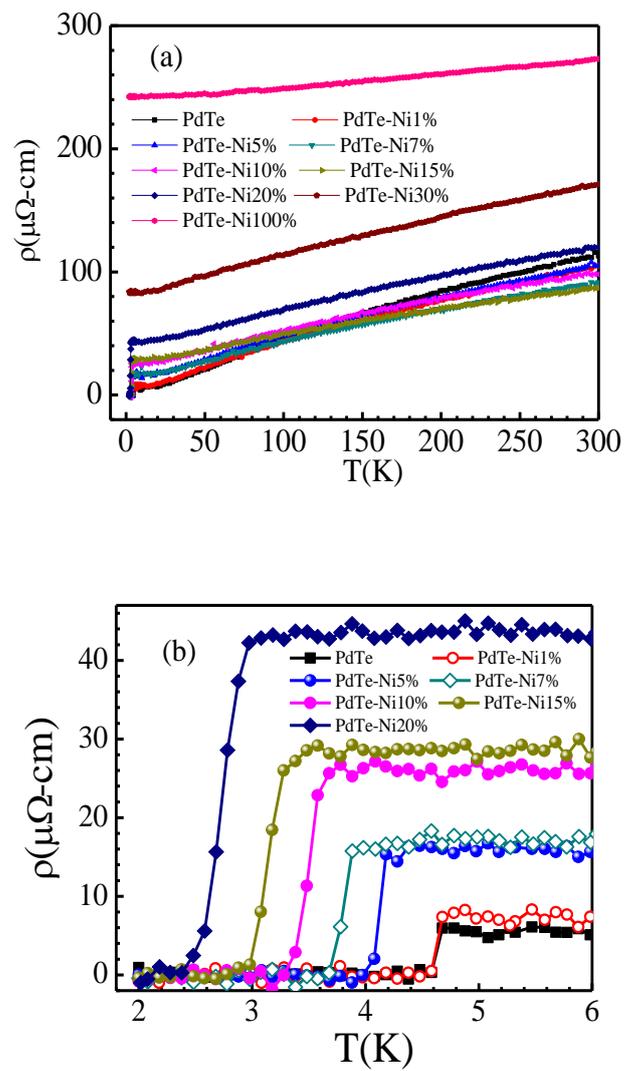

Fig. 6

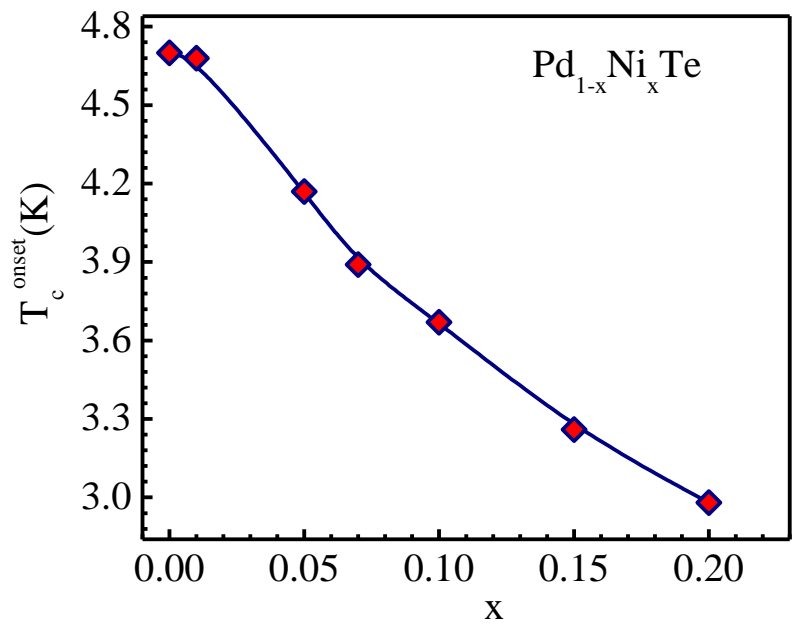

Fig 7:

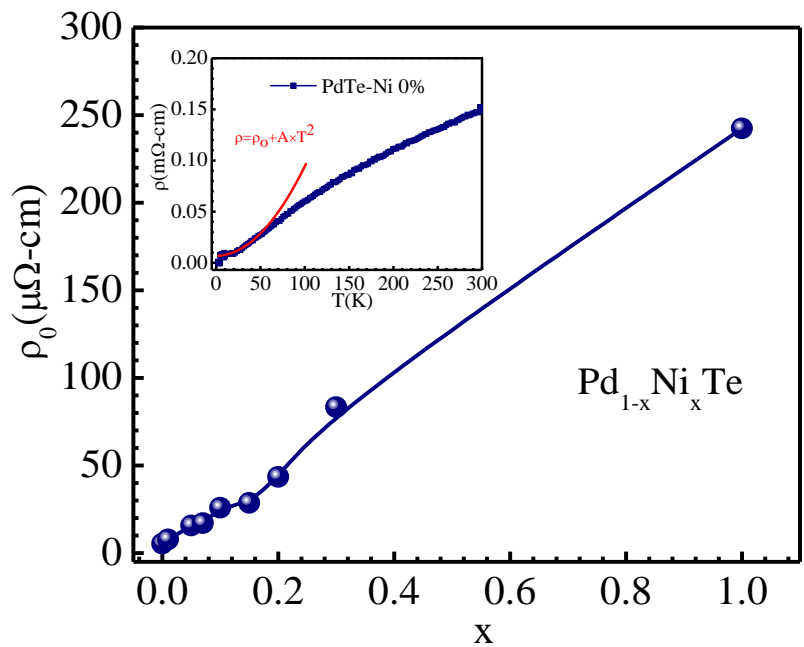

Fig.8:

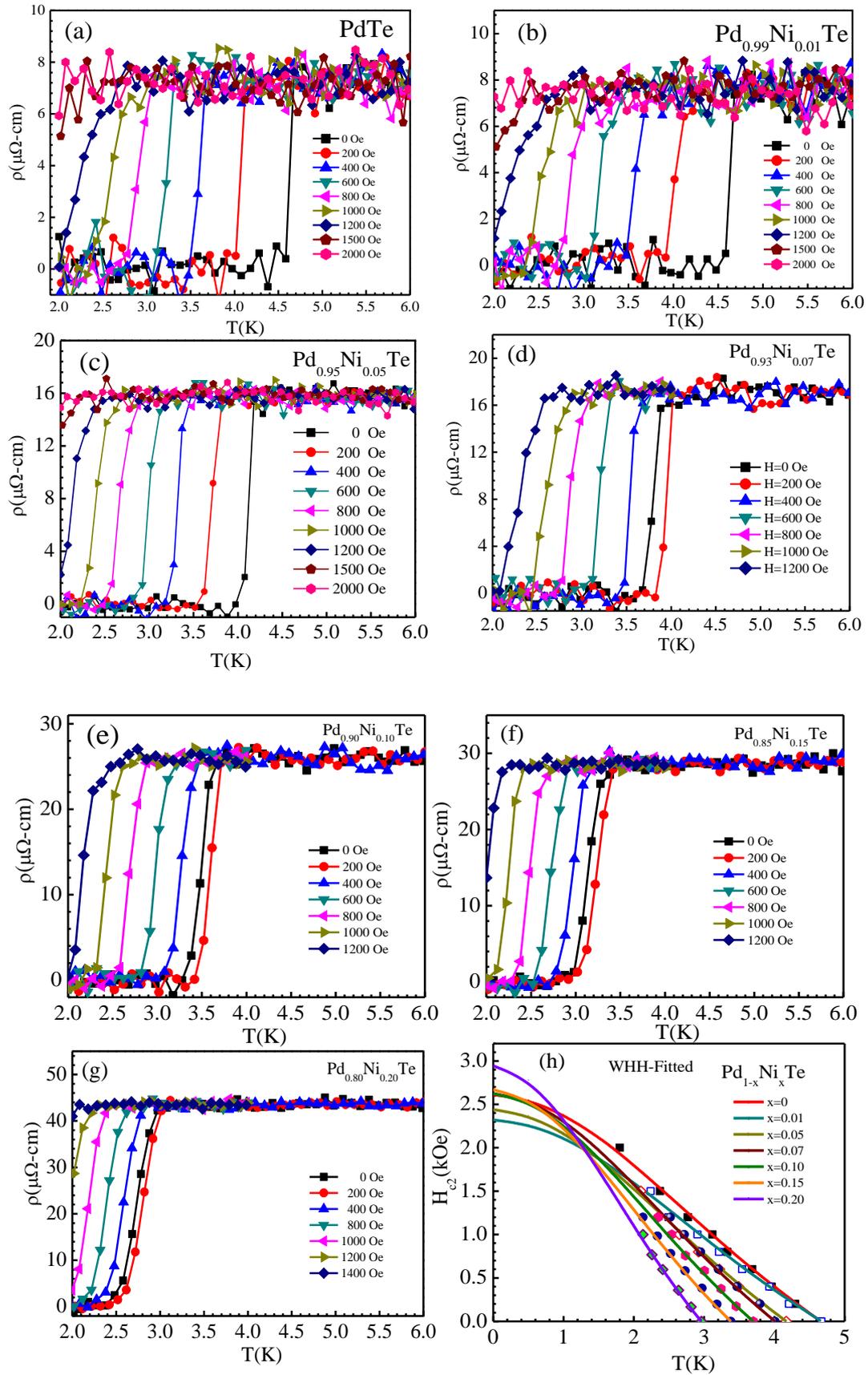

Fig 9:

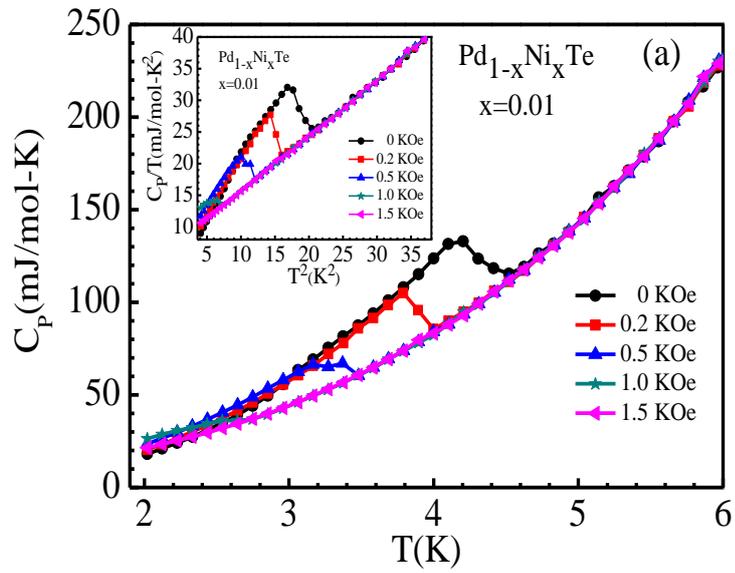

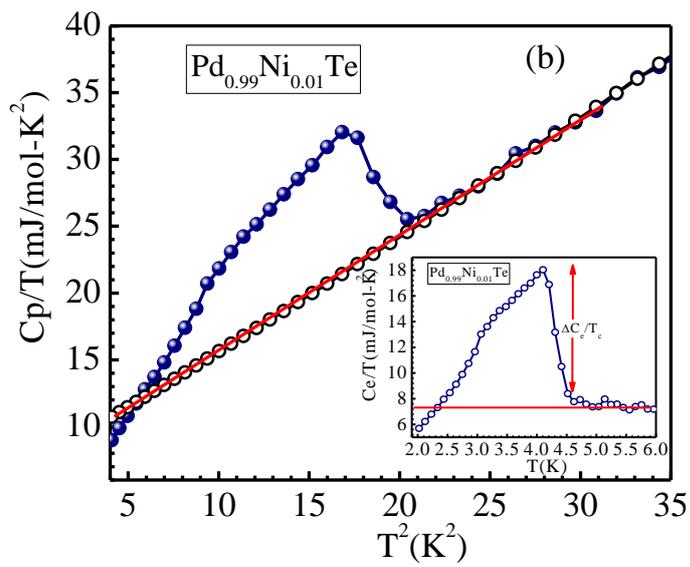